\documentclass{settings/WileyMSP-template}
\usepackage{tikz}

\usepackage[breaklinks]{hyperref}
\hypersetup{
    colorlinks=false,
    urlcolor=black,
    citecolor=black
    }
\def\checkmark{\tikz\fill[scale=0.4](0,.35) -- (.25,0) -- (1,.7) -- (.25,.15) -- cycle;} 

\usepackage{makecell}

\usepackage[utf8]{inputenc}
\usepackage{tabularray}

\usepackage{tablefootnote}

\usepackage{float}

\usepackage{ chemformula }

\begin{document}

\pagestyle{fancy}
\setlength\parindent{24pt}
\sloppy
\rhead{\includegraphics[width=2.5cm]{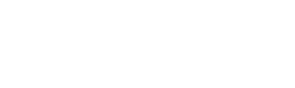}}
\title{Data-driven compound identification in atmospheric mass spectrometry} 

\maketitle


\author{Hilda Sandstr\"om}
\author{Matti Rissanen*}
\author{Juho Rousu}
\author{Patrick Rinke*}


\begin{affiliations}

Prof. M. Rissanen\\
Aerosol Physics
Laboratory \\
Tampere University\\
FI-33720, Tampere, Finland\\
Email Address: matti.rissanen@tuni.fi\\

Prof. J. Rousu\\
Department of Computer Science\\
Aalto University \\
P.O. Box 11000 \\
FI-00076 Aalto, Espoo, Finland\\

Prof. P. Rinke, Dr. H. Sandstr\"om\\
Department of Applied Physics \\
Aalto University \\
P.O. Box 11000 \\
FI-00076 Aalto, Espoo, Finland\\
Email Address: patrick.rinke@aalto.fi\\

Prof. M. Rissanen\\
Department of Chemistry \\
University of Helsinki\\
P.O. Box 55
A.I. Virtasen aukio 1\\
FI-00560, Helsinki, Finland\\
\end{affiliations}



\begin{abstract} Aerosol particles found in the atmosphere affect the climate and worsen air quality. To mitigate these adverse impacts, aerosol particle formation and aerosol chemistry in the atmosphere need to be better mapped out and understood. Currently, mass spectrometry is the single most important analytical technique in atmospheric chemistry and is used to track and identify compounds and processes. Large amounts of data are collected in each measurement of current time-of-flight and orbitrap mass spectrometers using modern rapid data acquisition practices. However, compound identification remains a major bottleneck during  data analysis due to lacking reference libraries and analysis tools. Data-driven compound identification approaches could alleviate the problem, yet remain rare to non-existent in atmospheric science. In this perspective, we review the current state of data-driven compound identification with mass spectrometry in atmospheric science, and discuss current challenges and possible future steps towards a digital era for atmospheric mass spectrometry.  
\end{abstract}

\section{Introduction}\label{sec:intro}
In this perspective article, we review the current state of data-driven mass spectrometry in atmospheric science. We focus on automated compound identification, which refers to the large-scale identification of molecules facilitated by digital tools, open knowledge and data sharing practices. The past 50 years have seen the emergence of large mass spectral databases, which are filled with mass spectra for a variety of compounds.\cite{Stein2012, E_etc_Stenhagen1969} Mass spectral databases are used during compound identification and the development of data-driven identification tools. As a result, many research fields, which rely on high-throughput mass spectrometry, have been able to improve, accelerate and automate data analysis of mass spectrometry experiments.  However, in atmospheric science, we believe that there is room for a broader application and more specific development of such tools. Here, we outline the potential and current barriers for data-driven compound identification in atmospheric mass spectrometry.

Atmospheric science includes the study of all chemical and physical processes that occur in the atmosphere. These processes drive a complex, interlinked system with global impact.  The chemical composition of the atmosphere mostly consists of nitrogen and oxygen gas (around 99\%), followed by noble gases (about 1\%), water vapor (ca. 0.01 to 4\%), and carbon dioxide (0.04\%). In addition, the atmospheric gas mixture contains a vast number of trace gases, including methane and carbon monoxide (around 2 ppm and 100 ppb, respectively), inorganic vapors, such as nitrogen and sulfur compounds (e.g., \ch{NO}, \ch{NO2} and \ch{HNO3}, and \ch{SO2}, \ch{COS} and \ch{CS2}), and a substantial number of organic compounds from either biogenic or anthropogenic emissions (e.g., terpenes and polyaromatics). These trace gases all transform in the atmosphere through reactions initiated by sunlight.\cite{Wayne_Richard_P_2000-03-30,Seinfeld_John_H_2006-08-11, FinlaysonPitts2000} 

\begin{figure}[H]
  \includegraphics[width=\columnwidth]{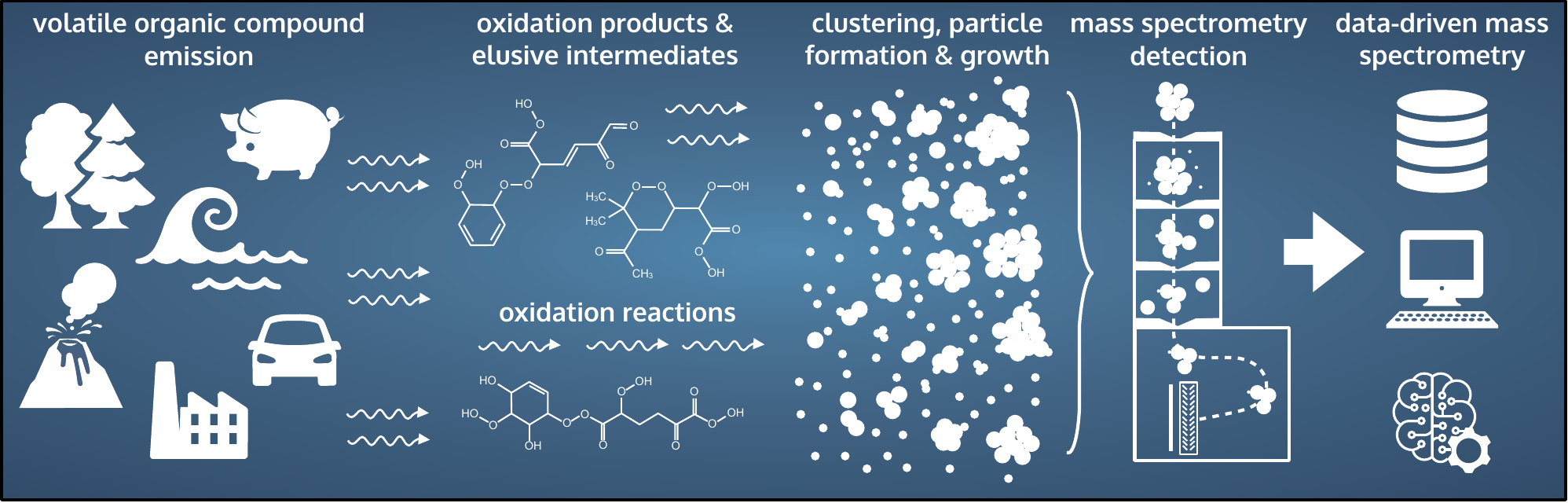}
  \caption{Particles in the atmosphere form through complex processes spanning multiple spatial-scales.  First, emissions of volatile compounds enter the atmosphere and oxidize into lower volatility compounds. These low-volatility compounds eventually form clusters which,  in turn, can grow into atmospheric nanoparticles. Mass spectrometry has become the measurement method of choice to study atmospheric molecular processes like these. Introducing data-driven methods such as machine learning to the mass spectrometry workflow can help unlock the full analytical potential of mass spectrometry, and provide unprecedented insight into atmospheric processes.}  
  \label{fig:intro}
\end{figure}

Trace gases can alter the atmospheric composition at any given time. Certain trace gases are very reactive and have short lifetimes, while others are practically non-reactive and persist for far longer periods, allowing them to transport over long distances. Trace gas emissions of organic compounds enter the atmosphere mainly in reduced and poorly water-soluble forms. Through oxidation, the organic compounds increase their affinity for the condensed phase (see\textbf{ Figure  \ref{fig:intro}}). This means they can be scavenged by liquid droplets and airborne particles. One example of this complex multi-phase chemistry is secondary organic aerosol particle generation. Secondary organic aerosol particles form via rapid gas-phase oxidation of emitted volatile organic compounds (also referred as VOCs) into low-volatile reaction products that can grow atmospheric aerosol particles,\cite{Kroll2008,Ehn2014, Donahue2012basis} or form them directly.\cite{Kirkby2016, Simon2020, Rose2018}  An autoxidation process drives this gas-to-particle conversion by generating a sequence of progressively more oxygenated, and often isomeric, reaction products from the same parent hydrocarbon.\cite{Crounse2013, Rissanen2014} With each oxygenation step the reactant molecules become better at condensing onto smaller nanoparticles.\cite{trostl16, Bianchi2019}  

The volatility of a compound and its tendency to form atmospheric secondary organic aerosol particles can be described conceptually by the volatility basis set.\cite{trostl16,Donahue2011,Schervish2020} The basis set contains information on the vapour concentration and oxygen content (the oxygen to carbon ratio, O:C, or the average carbon oxidation state, OSc) and correlates the volatility evolution with structural changes.  The most oxygenated, and generally also the most polar, compounds contribute most to aerosol particle formation and typically have the highest O:C ratios and lowest saturation vapour concentrations. The most extreme case are the so-called ultra low volatile organic compounds (ULVOCs) with saturation vapour concentrations lower than $3 \times 10^{-9}$ $\mu$g  m$^{-3}$.\cite{trostl16,Donahue2011,Schervish2020,Simon2020} At the opposite end of the volatility basis set scale, we find the most volatile, and the least polar, organic compound gases. 

The shear number of emitted volatile organic compounds, combined with the many plausible oxidation reaction schemes alluded to above, lead to a combinatorial explosion of possible reaction products.  The number of different, emitted volatile organic molecules is estimated to lie in the thousands or even millions.\cite{Goldstein2007, Noziere15} Through atmospheric reactions, each emitted volatile organic compound multiplies into thousands of reaction products. For example, a decane molecule (10-carbon alkane) with around 100 isomers, could already yield over one million distinct compounds.\cite{Goldstein2007}

Understanding the complex atmospheric chemistry behind aerosol particle formation is an important and challenging task. Efforts to map atmospheric compounds and processes contribute to a better basic knowledge of the chemistry in one of Earth's largest and most complex systems. The atmospheric chemistry leading to particle formation also contributes to air pollution and climate change. Aerosol particle pollution has adverse effects on air quality and human health,\cite{Pozzer2023} contributing to 7-9 million premature deaths annually.\cite{Khomenko2021, Lelieveld2020}  Additionally, aerosol particles impact the climate by  reflecting and absorbing solar radiation, an effect addressed in climate models used by the Intergovernmental Panel on Climate Change (IPCC) to inform and guide legislation and action plans for  climate change mitigation.\cite{IPCC2021} In this context, compound identification could, for example, help to develop a better understanding of particle growth, an important factor in determining aerosol-cloud interactions.\cite{ipcc_ts_21} Small changes in our understanding of aerosol particle growth could alter the number of cloud condensation nuclei by 50\%, and thus affect the outcome of climate models.\cite{trostl16}  In this perspective, we propose merging experimental mass spectrometry techniques with data-driven approaches, such as machine learning, to accelerate identification of new atmospheric compounds (see Figure  \ref{fig:intro}).  

\begin{figure}[H]
  \includegraphics{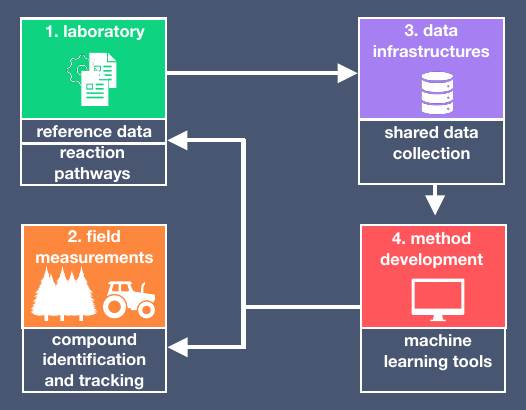}
  \caption{Data-driven compound identification in atmospheric mass spectrometry requires an integration of experiments and data science frameworks. Laboratory experiments can be used to create reference spectra for atmospheric compounds (1). Field measurements produce large amounts of mass spectrometry data of unknown compounds (2). Reference spectra and field measurements can be collected in shared data repositories (3). Data-driven (e.g. machine learning-based) compound identification tools can be trained with reference spectra and be used to identify new compounds measured in field campaigns or laboratories thereby increasing our basic knowledge of atmospheric processes (4).}  
  \label{fig:intro2}
\end{figure}

Atmospheric scientists utilize a combination of laboratory and field-campaign spectrometry experiments  to map out the intricacies of atmospheric chemistry leading to particle formation (\textbf{Figure \ref{fig:intro2}}).  Field-campaigns generate numerous experimental spectra of compound mixtures. Such mixtures often contain unknown compounds and have a composition that varies between measurement sites. Meanwhile, laboratory experiments can, for example, be used to create reference spectra to aid the identification and tracking of atmospheric compounds.\cite{Claeys2007,Parshintsev2008,Lin2012,eijck2013} In a data-driven approach, existing experimental infrastructures would be coupled to data science frameworks. Reference compounds shared in data infrastructures can function as training data for automated compound identification tools. Such digitization of atmospheric mass spectrometry could then expedite compound identification in laboratories and field measurements and help us to gain basic knowledge of the chemistry guiding particle formation (Figure \ref{fig:intro2}). 

\section{Mass spectrometry as a window into molecular-level atmospheric processes}  \label{sec:MSvol}
Much of what is currently known about atmospheric molecular-level processes was obtained with mass spectrometry. While mass spectrometers primarily provide data on the molecular mass and formula, the molecular formula alone often cannot uniquely identify a compound.\cite{Baker2019}  To gain additional insights into molecular structures,  mass spectrometry can be combined with techniques such as chromatographic separation,\cite{Grebe2011}  induced fragmentation (MS/MS\cite{Glish2003,Sadygov2004}  and electron ionization (EI) mass spectrometry,\cite{Yang2023EI}) ion mobility spectrometry,\cite{Skytta2022,Krechmer2016} ionization characteristics,\cite{Rose2021,Berresheim2000, Smith2010} and spectroscopy methods.\cite{Noziere15}  Such combined approaches have the potential to identify compounds and address a wide range of research questions, including those requiring high-throughput analysis. However, the use of mass spectrometry in atmospheric science faces many challenges, which we outline below. 

\begin{figure}[H]
  \includegraphics{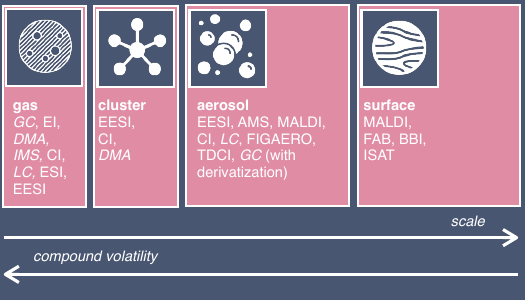}
  \caption{Example overview of mass spectrometric techniques, and complementary separation techniques (in italicized font), used to study atmospheric compounds ranging from molecules in the gas-phase, clusters to aerosols and aerosol surfaces. The arrows at the bottom of the figure indicate the inverse relation between measurable scale and detectable volatility. The acronyms in the figure are: EI - electron ionization; DMA - differential mobility analysis; IMS - ion mobility spectrometry; CI - chemical ionization; ESI - electrospray ionization; EESI - extractive electrospray ionization; AMS - aerosol mass spectrometry; MALDI - matrix assisted laser desorption ionization; FIGAERO - filter inlet for gas and aerosols ; TDCI - thermal desorption chemical ionization; FAB - fast atom bombardment; BBI - bursting bubble ionization; ISAT - interfacial sampling with an acoustic transducer.} 
   \label{fig:techniques}
\end{figure}

\textbf{Figure \ref{fig:techniques} }shows examples of mass spectrometric techniques used to study different compounds in atmospheric chemistry.\cite{Laskin2018} In the introduction, we alluded to the fact that atmospheric chemistry (gas, molecular clusters and particles) involves compounds with widely different volatility. Since mass spectrometry is inherently a gas-phase detection method any specimen must first be volatilized. For this purpose, specialized techniques have been developed to study low-volatile molecules with mass spectrometry.

 The experimentally resolvable fraction of compounds, in terms of their volatility, has expanded steadily, as techniques have improved.\cite{Glish2003,Tamara2022} For example, large biomolecules have been detected using several spray ionization sources (e.g., electrospray ionization, ESI,\cite{Wilm2011, Fenn1989} and atmospheric pressure photoionization, APPI),\cite{Gaudin2012,Bagag2012,Sedlackova2022} and surface-bound species by desorption techniques such as matrix assisted laser desorption ionization (MALDI).\cite{Fei1996,Glish2003} Particulate bound targets, the constituents of nanoparticles, can be detected through direct aerosol sampling by e.g., using an aerodynamic lens with subsequent flash vaporization and EI ionization in aerosol mass spectrometry (AMS),\cite{Jayne2000} or by collecting the particles onto a filter (or wire) with subsequent rapid thermal desorption vaporization of the condensed-phase constituents. The latter is, for example, applied in chemical ionization mass spectrometry (CIMS)\cite{Munson1966,Munson1977} detection (with, e.g., filter inlets for gas and aerosols, FIGAERO,\cite{LopezHilfiker14} or thermal desorption multi-scheme chemical ionization inlet, TD-MION,\cite{Partovi2023}). 

 Of the atmospheric compounds, the volatile gas-phase organic molecules are commonly investigated with either gas-chromatography mass spectrometry (GC-MS)\cite{Bartle2002} or proton transfer reaction mass spectrometry (PTRMS).\cite{Lindinger1998} The least volatile fraction (corresponding to the lowest gas-phase concentrations) can generally only be measured  by atmospheric pressure interface (Api) CIMS methods employing anion attachment.\cite{Jokinen2012,Ehn2014,Simon2020} Finding techniques that are applicable to the whole range of molecular species present in the atmosphere is a major challenge in atmospheric mass spectrometry, and multiple techniques are currently required to cover the whole volatility range (Figure \ref{fig:techniques}). 

Besides a broad compound coverage, the ideal mass spectrometric technique in atmospheric science should be able to analyse ambient gas-phase samples directly without the need for sample pre-treatment.\cite{Rissanen2019} However, such techniques  are rare, and are often limited by, for example, sampling requirements (e.g., limited time resolution resulting from the necessary temporal spacing of compounds as they pass through a chromatographic column), sensitivity, and interference from background compounds (e.g., spectral overlaps in spectroscopic techniques).\cite{Rissanen2021,Tiusanen2023}  Api-CIMS is popular, because it can sample ambient air, usually through a differentially pumped interface (see e.g., ref \cite{Junninen2010}). Samples do not need to be pre-treated, which enables direct, online analysis. While various methods exist for analyzing aerosols in real-time, such as resonance multiphoton ionization (REMPI)\cite{Adam2011, Laskin2012} and  secondary electrospray ionization (SESI),\cite{Semren2022} we will focus here on Api-CIMS due to its user-friendliness, reliability, and robustness. Api-CIMS can operate continuously for months, even in field conditions. Without sample pre-treatment, Api-CIMS can be coupled with other research methodologies, which provide complimentary information, such as ion mobility.\cite{Krechmer2016,Skytta2022}  Api-CIMS is most commonly applied in ambient field measurements and environmental chamber campaigns where it is combined with several other measurement techniques.\cite{Ehn2014, Almeida2013,Sipila16,Mauldin2012, Wang2020, Eisele1993}

The atmospheric composition at a research site can be monitored for days, weeks, or sometimes even years. These time-consuming field campaigns are characteristic of atmospheric mass spectrometry, and set atmospheric science apart from other research fields that use mass spectrometry (e.g., metabolomics or pharmaceutics).\cite{Mayr2009}  Field instruments usually produce relatively long time series for a selected group of target ion signals.\cite{Rose2021,Berresheim2000} At the opposite end of the time spectrum, specimen can also be collected on a filter or a filament and then analyzed within a few minutes in an Api-CIMS\cite{Smith2010,LopezHilfiker14,Partovi2023} enabling high-throughput studies of e.g., aerosol particles. While early quadrupole-based Api-CIMS instruments were by necessity only monitoring selected target ions, modern mass spectrometric methods measure the whole mass spectrum continuously.\cite{Glish2003} The field measurements are often performed up to a mass resolution of 200 000 (the higher the mass resolution, the smaller the resolvable changes in the target mass), which generates large amounts of data  that make data analysis challenging. 

Currently, only a fraction of compounds in atmospheric mass spectrometry measurements are definitively identified due to the various challenges we will review in the next section.\cite{Noziere15} Two possible mass spectrometry approaches exist that are suitable for compound identification following or during field campaigns. For example, compounds collected on-site can be analyzed later in the laboratory with chromatography and fragmentation mass spectrometry.\cite{Worton2017,Franklin2022,Hamilton2004} Alternatively, current developments for improved compound identification by other mass spectrometry techniques used during field-campaigns are ongoing, and outlined below. 

Field campaigns often employ soft ionization approaches such as Api-CIMS, which minimize ion fragmentation. In Api-CIMS, reagent ions attach to target molecules (adduction mode), revealing molecular formula information. Details on the molecular structure can be obtained by coupling Api-CIMS with molecular fragmentation techniques (MS/MS).\cite{Tomaz2021} Varying the reagent ion increases sensitivity and selectivity, with detectable target ion concentrations ranging down to $10^{-4}$ cm${^{-3}}$.\cite{Jokinen2012,Bianchi2019,Hyttinen2018,Iyer2017}  New methods, e.g., selected ion flow tube mass spectrometry (SIFT-MS) and specialized CIMS,\cite{Brophy2015} have been developed to improve compound identification by varying the ion-molecule interaction. Noteworthy is the 2019 development of the MION inlet platform,\cite{Rissanen2019} facilitating rapid transitions between ionization modes (e.g., nitrate in anion mode\cite{Hyttinen2015}  and aminium- or proton-transfer in the cation mode\cite{Berndt2018}). MION has already increased the number of detectable  atmospheric molecules\cite{Rissanen2019,He2023} and further methodological synergy promises even better compound identification in atmospheric mass spectrometry.\cite{Iyer2016,Hyttinen2018} 

Summarizing this section, atmospheric science is in a state of dichotomy. Field campaigns have produced large amounts of data, but this data is not labeled and has not been uploaded to mass spectral databases (see following sections). Moreover, development of data-driven compound identification tools, and the accuracy of the tools after deployment, relies on the production and analysis of coherent high-quality reference data.\cite{Hamilton2004,Worton2017,Franklin2022}   The vast atmospheric compound space, the heterogeneity of studies (field- vs laboratory), and the multiple mass spectrometric techniques have produced a data landscape that is difficult to navigate. Standardisation procedures for data collection, processing and analysis are still lacking. Combined, these challenges have aggravated compound identification in atmospheric science. 

\section{Compound identification with mass spectrometry}\label{sec:CI}
The identification of unknown compounds and processes is the holy grail of atmospheric mass spectrometry. To identify unknown processes and compounds is challenging, requiring  suitable identification techniques and  a high accuracy identification method. Since only a few hundred atmospheric compounds out of potentially millions have been identified in aerosol samples,\cite{Hamilton2004,Worton2017,Franklin2022}  the chemical space of atmospheric compounds remains largely uncharted. We also note that, while compound identification is important for gaining basic knowledge of atmospheric chemistry and for use in particle formation modeling,\cite{Elm2020}  atmospheric mass spectrometry studies are diverse in type and aim. Some studies do not require compound identification, such as: I) inventorying compounds based on their properties II) real-time monitoring, or  III) monitoring known sources or processes (for a review, see ref \cite{Noziere15}). In these example cases, it can be sufficient to track a molecular or elemental composition, or specific compounds and sources, which are easier objectives than compound identification. 

In this perspective, we focus on compound identification. We have identified three factors that most affect the accuracy of compound identification in mass spectrometry that we will present in more detail in the following: the chosen experimental technique, the compound identification method (or tool) and the existence of reference standards. 

Mass spectrometry methods are able to identify compounds to a varying degree. In 2015, Nozi\`ere et al. introduced the I-factor to quantify the identification accuracy of a mass spectrometry technique in terms of the  ability to narrow down the number of plausible candidate structures.\cite{Noziere15} In the best case, only one plausible structure is identified and the I-factor is equal to one. If the identification method is not able to discern between isomers of the molecular formula, the I-factor goes up to the number of isomers (two or higher). Uncertainties in the determination of the molecular formula can further increase the I-factor. 

Nozi\`ere et al. used the I-factor to compare atmospheric mass spectrometric techniques in terms of their compound identification ability.\cite{Noziere15} The best I-factors were achieved when two or more techniques, such as chromatography and mass spectrometry, were combined. Fragmentation mass spectrometry methods such as tandem mass spectrometry and EI mass spectrometry, coupled to chromatography methods, reached I-factors of 1-3. The I-factor of soft ionization techniques like CIMS were estimated around 4-40 at the time of publication. The newly developed  MION-CIMS method, that uses multiple ion chemistries (see Section \ref{sec:MSvol}), has the potential to achieve similarly low I-factors as the combination of two or more techniques given above.\cite{Rissanen2019,Rissanen2021} The data produced by mass spectrometry techniques is used to isolate candidate structures with the help of a compound identification method. 

The identification accuracy of compound identification methods and tools varies and is determined by their ability to match a recorded spectrum to a molecular structure. In Section \ref{sec:tools} we summarize these tools and their principles. The performance of a compound identification tool is measured by the \textit{Top-k} accuracy. Unlike the I-factor, which quantifies the ability of a mass spectrometry technique to resolve the identity of a compound, the \textit{Top-k} accuracy gives the percentage of instances in which the correct compound is found among the \textit{k} best matching compounds during a compound search.  For example, a benchmark study in ref \cite{Durkop2019} reported a Top-1 accuracy of 39.4 (and a Top-10 accuracy of 74.8) for their highest-ranking identification tool. This means that the tool identified the correct molecular structure in two out of five cases (Top-1 accuracy of 39.4) and found it among the ten best matches in three fourths of all cases (Top-10 accuracy of 74.8). Here it should be noted that the absolute numbers are highly dependent on both the data size used in training and the molecular database used to retrieve candidate molecular structures. Moreover, the recorded mass spectrum's quality and type can limit the compound identification method's ability to provide reasonable candidate structure suggestions. 

The accuracy of a compound identification tool often depends on the existence of appropriate reference standards, i.e., measured mass spectra of compounds, which are either identical or similar to the unknown compound.  In the compound identification process, most approaches search for the measured spectrum, or a very similar one, in a database. Even if the identification method does not employ a spectral database search, it has still likely been developed, parameterized or trained with data from one or more such databases.  In atmospheric science, the lack of reference standards is a large barrier for effective compound identification,\cite{Noziere15,Bianchi2019,Rissanen2021} which we will return to later in this perspective.

\begin{figure}[H]
  \includegraphics{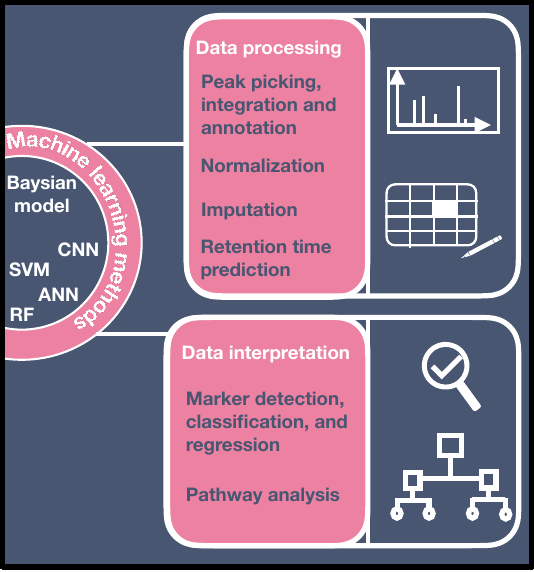}
  \caption{Data processing and analysis steps in a mass spectrometry experiment which have been performed using machine learning methods. Spectral information is extracted through data processing and analysis. Data processing serves to mitigate statistical effects such as batch-to-batch variations, or missing data. Other processing steps include peak processing, alignment, integration, and annotation.  Conversely, data analysis aids in the classification or detection of molecules, and the identification of chemical pathways to the observed molecules. Acronyms: ANN - artificial neural network; CNN - convolutional neural network; RF - random forest model; SVM - support vector machine.} 
  \label{fig:workflow}
\end{figure}

In the digitization of compound identification  in atmospheric mass spectrometry, machine learning will naturally play a large role. As we will detail in the next section, machine learning tools are already utilized to automate and improve analysis and processing of mass spectrometry data in other fields (see a recent review in ref \cite{Liebal2020}). \textbf{Figure \ref{fig:workflow}} illustrates a typical mass spectrometry data acquisition process.  In atmospheric mass spectrometry, machine learning is already applied to some, but not all, of the steps outlined in Figure \ref{fig:workflow}. Machine learning models have been trained on different atmospheric mass spectrometry data (like AMS,  PTRMS, ESI-mass spectrometry, single particle mass spectrometry, and inductively coupled plasma mass spectrometry) for aerosol classification and source apportionment,\cite{Phares2001,Murphy2003,Aijala2019,Zawadowicz2017,Christopoulos2018,Lu2022,Wang2021rev,Bland2022,Gong2022} prediction of composition\cite{Giri2021,Pande2022,Zhang2022PM,Feng2022} and properties.\cite{RuizJimenez2021,Jiang2021} Moreover, a recent review highlighted the role of machine learning in data pre-processing during measurements of volatile organic compounds.\cite{Sun2018} Thus, machine learning is being integrated into the data analysis of atmospheric mass spectrometry, but little attention is currently devoted to compound identification. A GC-MS machine learning model for molecular formula annotation of atmospheric, halogenated compounds,\cite{Guillevic2021} or for molecular property and quantification factor prediction,\cite{Franklin2022} are two notable exceptions.

We will next address the reasons for the gap between the perceived demand and utility of smart, high-throughput compound identification tools for atmospheric mass spectrometry and the lack of corresponding availability of such tools. We will also identify the major barriers for introducing compound identification techniques in atmospheric mass spectrometry. A key to both these points are  currently available mass spectral databases and their link to the success story of machine learning for compound identification in the field of metabolomics.

\section{Mass spectral databases}\label{sec:mb_tools}
Digital mass spectrometry libraries with reference mass spectra, so called mass spectral databases, have been used for compound identification since the 1960s.\cite{Stein2012, E_etc_Stenhagen1969}  Over time, mass spectral databases have grown in size and usage, partly as a result of increased data processing and storage capabilities as well as adoption of open science practices. Table \ref{grid_mlmmh} summarizes a selection of mass spectral databases that are hosted by research institutions, or distributed by companies and mass spectrometry vendors.  The mass spectral data is either collected through research community contributions (e.g., refs \cite{Wang2016,Hummel2013,Sud2007, MassBankEu22,MONA23,Sawada2012,mzCloud}), or curation of scientific publications, measurements and computations (e.g., refs \cite{Wishart2022,Watanabe2000,Kind2013,Weber2012,Wallace2023,Taguchi2010,Wissenbach2011_dev,Wissenbach2011_drugs}) .

\DefTblrTemplate{capcont}{hilda}{%
    \DefTblrTemplate{caption-sep}{hilda}{ - }
    \DefTblrTemplate{conthead-text}{hilda}{continued from previous page}
    \DefTblrTemplate{caption-tag}{hilda}{Table\hspace{0.25em}\thetable}
    \par\centering
        \UseTblrTemplate {caption-tag}{hilda}
        \UseTblrTemplate {caption-sep}{hilda}
        \UseTblrTemplate {conthead-text}{hilda}
    \par
}
\NewTblrTheme{hildacaption}{%
    \SetTblrTemplate{capcont}{hilda}%
    
}

\begin{small}
\begin{longtblr}[theme = hildacaption,
    caption = {List of select mass spectrometry databases. The list is divided into open access (top) and commercial (bottom). Data volumes reflect the state in August 2023 (The data was taken from an associated webpage or publication). Acronyms: GC - gas chromatography; MS - mass spectrometry; FAB - fast atom bombardment; MS/MS - tandem MS; LC - liquid chromatography; MS$^n$ - tandem mass spectrometry done with n fragmentation stages.},
    label = {grid_mlmmh},
]{
    colspec = {X[0.225,j] X[0.2,j] X[0.475,j] X[0.1,j]}, 
    colsep=5pt, 
    width = 1.0\textwidth,
    rowhead = 1,
    row{1} = {font=\centering\itshape},
}

    \hline
    Name & Website & Description  & Reference \\\hline

    Global Natural Product  Social Molecular Networking (GNPS) & \href{https://gnps.ucsd.edu}{gnps.ucsd.edu} & The database contains 26485 unique structures (when full structure is available). The GNPS database contains data contributions from the public and other mass spectral libraries. & \cite{Wang2016}  \\
    
    Golm Metabolome Database & \href{http://gmd.mpimp-golm.mpg.de}{gmd.mpimp-golm.mpg.de}  & Public database maintained by the Max Planck Institute of Molecular Plant Physiology containing 26590 mass spectra. Has GC-MS spectra for 2222 metabolites and 3651 reference substances. & \cite{Hummel2013}\\
    
    Human Metabolome  Database (HMDB), v5 & \href{https://hmdb.ca}{hmdb.ca} & Freely available database containing experimental and predicted mass spectra. The database has predicted and experimental GC-MS spectra for 74944 and 3000 compounds, respectively, as well as predicted and experimental LC-MS/MS spectra for 206809 and 4064 compounds, respectively. HMDB also contains predicted retention times and collision cross sections. & \cite{Wishart2022} \\
    
    LipidBank & \href{http:\\lipidbank.jp}{lipidbank.jp}  & Curated database containing $>$ 6000 lipids and their spectral information (EI-MS, FAB-MS), & \cite{Watanabe2000}\\
    
    LipidBlast & \href{https://fiehnlab.ucdavis.edu}{fiehnlab.ucdavis.edu} & An in silico tandem mass spectral library for lipid identification containing predicted spectra for 119200 compounds. Provides a tool for users to predict new spectra for their molecules, available in MS-Dial software. & \cite{Kind2013} \\

    Lipid Maps  Structure Database (LMSD) & \href{www.lipidmaps.org}{lipidmaps.org} & LMSD is a database of $>$ 48169 lipid structures, 26122 of which were determined experimentally and 22047 of which were generated computationally. LMSD has links to in-house  (500 lipid standards) and external (54877 MS and MS/MS spectra for 7210 lipids from MassBank of North America) mass spectrometry resources.  & \cite{Sud2007} \\
    
    MaConDa, v1 & \href{www.maconda.bham.ac.uk}{maconda.bham.ac.uk}  & Freely available, manually annotated database of ~200 known small molecule contaminants and their LC-MS and GC-MS peaks. Contains un-annotated data. Downloadable and searchable in batch format. & \cite{Weber2012} \\
    
    MassBank (EU), v2023.09 & \href{https://massbank.eu}{massbank.eu}  & Public repository of $>$96449 mass spectra of $\geq$ 15500 molecules in metabolomics, exposomics and  environmental samples. & \cite{MassBankEu22} \\
    
    MassBank of North America (MoNA) & \href{https://mona.fiehnlab.ucdavis.edu}{mona.fiehnlab.ucdavis.edu}  & Auto-curated public database with experimental and computational  mass spectra  of $>$ 650292 compounds.  Includes quality estimation of the mass spectra. & \cite{MONA23} \\
    
    Advanced Mass Spectral Database (mzCloud) & \href{https://beta.mzcloud.org}{beta.mzcloud.org} & Commerical database maintained by HighChem LLC,  Slovakia with manually curated high-resolution LC-MS/MS spectra for 26417 compounds. & \cite{mzCloud}  \\
    
    RIKEN tandem mass spectral database (ReSpect) for phytochemicals & \href{http://spectra.psc.riken.jp}{spectra.psc.riken.jp}  & A curated database with 8649 tandem mass spectra of 3595 plant metabolite compounds collected from scientific literature in 2011 and authentic standards. Has grown since and now contains 9017 (+368) spectra. & \cite{Sawada2012}  \\    \hline
    
    Maurer/Wissenbach/Weber  LC-MS$^n$ Library of Drugs, Poisons,  and
their Metabolites, (2nd edition) & \href{www.sciencesolutions.wiley.com}{sciencesolutions.wiley.com}  & LC-MS$^n$ library of over 2270 compounds and over 3600 of their metabolites curated for forensic use.& \cite{Wissenbach2011_dev,Wissenbach2011_drugs} \\

    Metlin Gen2  (Mass consortium) & \href{www.massconsortium.com}{massconsortium.com} & METLIN is a highly curated commercial database with experimental spectra on over 930,000 molecular standards (2023) (LC-MS/MS). All molecular standards were analyzed in positive and negative ionization modes and at four different collision energies (0 eV, 10 eV, 20 eV, and 40 eV).  & \cite{Smith2005,Guijas2018, Montenegro2020} \\
    
    NIST Tandem and Electron Ionization Mass spectral library, 2023 release & \href{https://chemdata.nist.gov/}{chemdata.nist.gov}  & Curated spectra of 51501 compounds (tandem) and 347,100 (EI), mainly metabolites, drugs, pesticides, peptides and lipids. Also contains a rentention index database, including predicted values. & \cite{Wallace2023} \\
    
    LipidSearch  (Thermofisher) & \href{www.thermofisher.com}{thermofisher.com}  & Computational database containing in-silico LC-MS  and LC-MS/MS spectra for $>$ 1.7 million lipid compounds. & \cite{Taguchi2010} \\
    
    Wiley Registry\textsuperscript{\textregistered} of Mass Spectral Data 2023 & \href{www.sciencesolutions.wiley.com}{sciencesolutions.wiley.com}   & A curated GC-MS library with 873000 spectra of 741000 unique compounds with relevance to applications in environmental, forensics/
toxicology, metabolomics, pharmaceutical, biotech, food/cosmetics, defense/
homeland security, and more. &  \cite{Wiley12}\\

    Wiley Registry\textsuperscript{\textregistered} of Tandem Mass Spectral Data - MS for ID & \href{www.sciencesolutions.wiley.com}{sciencesolutions.wiley.com}   & A curated LC-MS/MS library with spectra for 1163  compounds including illicit drugs, pharmaceutical  compounds, pesticides, and other small bioorganic molecules. & \cite{wileyTandem}\\
     \hline
\end{longtblr}
\end{small}

\begin{figure}[H]
  \includegraphics{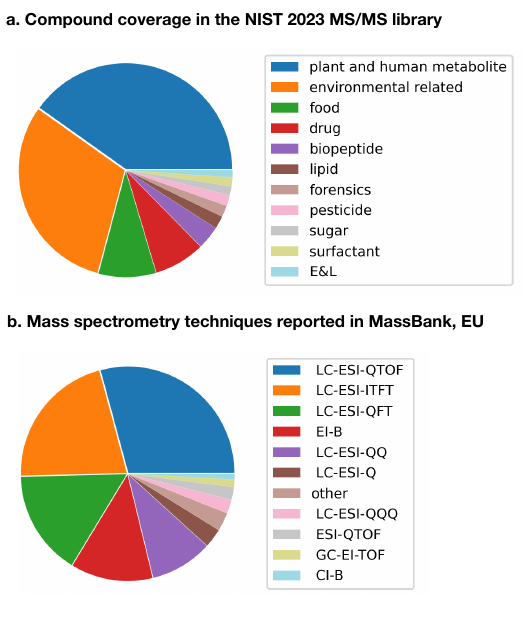}
  \caption{Example of listed contents in mass spectral databases. a. The reported compound coverage of the NIST 23 tandem mass spectral library. b. The different reported mass spectrometric techniques in the European MassBank. These two databases represent general mass spectral databases. Acronyms: E$\&$L - extractables and leachables; CI - chemical ionization ; B - bombardment; GC- gas chromatography; EI -  electron ionization; TOF - time-of-flight; ESI - electrospray ionization; Q, QQ, QQQ - single, double, triple quadrupole instrument; LC - liquid chromatography;  EI-B - electron bombardment ionization; QFT - quadrupole Fourier transform; ITFT -  inductively coupled plasma Fourier transform.}
  \label{fig:pies_on_massbanks}
\end{figure}

By design, mass spectral databases  either cover a specific compound space or aim for some level of generality. However, in reality, the data in large mass spectral databases tends to reflect the interest of the primary users and contributors. This is evident in \textbf{Table \ref{grid_mlmmh}}, which includes specific mass spectral databases created  for and by the metabolomics community. These databases contain predominantly small molecules called metabolites, found in organisms, cells or tissues. As in atmospheric science, mass spectrometry is used in metabolomics to identify and quantify molecules of interest. The plethora of mass spectral databases in metabolomics can be attributed to open science initiatives in the research field and the ensuing rapid growth over the past 25 years. As a result, large, general mass spectral databases contain mostly metabolites (see also \textbf{Figure \ref{fig:pies_on_massbanks}a}),\cite{Wallace2023,MassBankEu22,MONA23} despite no stated limitation or constraints on the compound coverage. For this reason, we have decided to  highlight metabolomics in this perspective and to use it as a comparative example for developments in atmospheric science. Besides metabolites, other common compound classes in general databases include molecules found in drug- or environmental samples (see an overview of NIST 2023 tandem mass spectral library in Figure \ref{fig:pies_on_massbanks}a). 

Mass spectral databases provide data collected with a variety of mass spectrometric techniques.  As can be seen in Table \ref{grid_mlmmh}, some databases focus on only one technique, such as LC-MS/MS,\cite{Kind2013,Sud2007,mzCloud,Sawada2012,Wissenbach2011_dev,Wissenbach2011_drugs,Montenegro2020,Taguchi2010,wileyTandem} or GC-MS,\cite{Hummel2013,Watanabe2000,Wiley12} while others provide data from  two or more techniques.\cite{Wang2016,Wishart2022,Wallace2023,Weber2012, MassBankEu22,MONA23} The most common technique is LC-MS/MS mass spectrometry followed by GC-MS.
For example, the MassBank of North America contains approximately 30 times fewer MS1 spectra (22500) than tandem mass spectra (including all MS$^n$) (May, 2023).  As expected, these most common mass spectrometric techniques found in mass spectral databases are those that facilitate compound identification (see Section \ref{sec:CI}). 

The number of compounds in the mass spectral databases of Table \ref{grid_mlmmh} varies considerably, although a direct comparison of the  database size is complicated by the non-standardized way in which the size is reported  (e.g., number of ions, or number of unique compounds, or number of spectra).  The reported data volume of mass spectral libraries either increases continuously or with new versions. The data volumes listed in Table  \ref{grid_mlmmh} reflect the state in August 2023. LipidSearch by Thermofisher is the largest mass spectral database with spectra for over 1.7 million lipid ions. Massbank of North America is the largest open access database with spectra for over 650000 compounds. The smallest database reports spectra for only 200 compounds .\cite{Weber2012} The median size of all databases reported in Table \ref{grid_mlmmh} is 26485 (average $>$ 290000). However, the databases overlap in terms of the compounds they cover.\cite{Vinaixa2016} The total amount of compounds offered by all databases together is therefore likely less than the sum of their individual compound counts.

Synthetic (i.e. computational) mass spectra have been important for creating large mass spectral databases. Table \ref{grid_mlmmh} also lists mass spectral libraries with computationally predicted (so called in silico) tandem mass spectra or GC-MS spectra.\cite{Sud2007,Kind2013,Taguchi2010,Montenegro2020,MONA23,Wishart2022} For example, LipidBlast is a purely computational  database,  which also provides a tool for users to build their own tandem mass spectrometry database.\cite{Kind2013}  The motivation for generating computational databases, and sometimes combining them with experimental ones, is the need to accelerate data collection. The large number of predicted mass spectra can greatly increase the average mass spectral database size. For example, HMDB contains experimental LC-MS/MS spectra for approximately 4000 compounds, but computational spectra for more than 200000 compounds. The quality and information content of in silico spectra is, however, still a subject of debate.  

The retention time provides useful additional information and is often enough for correct compound annotation in LC- and GC- mass spectrometry. However, for certain isomeric compounds even the simple chromatographic separation does not provide a positive compound identification and further separation can be necessary.\cite{Eckberg2019} Retention times in GC-MS are collected in MassBanks,\cite{MassBankEu22,Horai2010,MONA23} GMD,\cite{Hummel2013} and NIST23,\cite{Wallace2023} among others. In addition, computationally predicted retention times are supplied in, e.g., HMDB.\cite{Wishart2022} However, retention times tend to vary significantly between laboratories which hampers their utility for compound identification. Machine learning techniques can help in alleviating this problem (see Section \ref{sec:tools}).

Vinaixa and colleagues  have reviewed features of mass spectral databases in 2016.\cite{Vinaixa2016} They identified beneficial features such as open access, downloadable, large size, curation, data from different platforms, functionality to merge spectra, inclusion of chemical standards and addition of unknown compounds. On the adverse side, they list commercial licenses, lack of curation and spectrum information, limited sample sources, only negative polarity mode, or only computational data.  The review also surmises that there might be a trade-off between too many and too few instrument types as well as collision energies. Following Vinaixa et al., we summarize some features of the  mass spectral databases in Table \ref{grid_mlmmh} and \textbf{Table \ref{table_checklist}}.    
 
\begingroup 
\small
\centering

\begin{longtblr}[theme = hildacaption,
    caption = {Features of the mass spectrometry databases. Open access - partial of full free access to mass spectral data.  Data upload - users can contribute with data. Comp. data - Contains computationally (in silico) generated mass spectra. Exp. data - Experimental mass spectrometry data. Collects unknowns - Collects and adds unknown spectral queries. Machine learning tools - has associated machine learning tools.},
    entry = {Features of the mass spectrometry databases. Open access - partial of full free access to mass spectral data. Data upload - anyone can contribute. Comp. data - Contains computationally (in silico) generated mass spectra. Exp. data - Experimental mass spectrometry data. Collects unknowns - Collects and adds unknown spectral queries. Machine learning tools - has associated machine learning tools.},
    label = {table_checklist},
]{
    colspec = {X[0.3,j] X[0.1125,j] X[0.075,j] X[0.1125,j] X[0.1125,j] X[0.175,j] X[0.1125,j]}, 
    colsep=5pt, 
    width = 1.0\textwidth,
    rowhead = 1,
    row{1} = {font=\itshape}, 
}
    \hline         
    & Open access & Data upload & Computational data & Experimental data & Collects unknowns & Machine learning tools \\ \hline 
    
    Global Natural Products  Social Molecular Networking (GNPS) & \checkmark  & \checkmark & &\checkmark & \checkmark & \checkmark \\    
    
    Golm Metabolome Database & \checkmark\textsuperscript{\hyperref[fn:tab2_one_bottom]{1}\protect\phantomsection\label{fn:tab2_one_top}}   &  & \textsuperscript{\hyperref[fn:tab2_two_bottom]{2}\protect\phantomsection\label{fn:tab2_two_top}}  &\checkmark & \checkmark & \checkmark \\
    
    Human Metabolome  Database (HMDB), v5  & \checkmark & & \checkmark &\checkmark & \checkmark & \checkmark \\
    
    LipidBank & \checkmark  &  & & \checkmark & &\\
    
    LipidBlast  & \checkmark & \checkmark\textsuperscript{\hyperref[fn:tab2_three_bottom]{3}\protect\phantomsection\label{fn:tab2_three_top}} &\checkmark  & & &\\
    
    Lipid Maps Structure Database (LMSD) & \checkmark & \checkmark & \checkmark & \checkmark & & \\
    
    MaConDa, v1  & \checkmark & & \checkmark  &\checkmark & &\\
    
    MassBank (EU), v2023.09 & \checkmark & \checkmark &  &\checkmark & \checkmark\textsuperscript{\hyperref[fn:tab2_four_bottom]{4}\protect\phantomsection\label{fn:tab2_four_top}} & \\
    
    MassBank of North America (MoNA)  & \checkmark & \checkmark & \checkmark &\checkmark & & \\
    
    Advanced Mass Spectral Database (mzCloud)   & \checkmark & & &\checkmark & &\\ 
    
    RIKEN tandem mass spectral database (ReSpect) for phytochemicals & \checkmark &  & & \checkmark& & \\  

    Maurer/Wissenbach/Weber  LC-MS$^n$ Library of Drugs, Poisons,  and their Metabolites, (2nd edition) & & & & \checkmark & &\\

    Metlin Gen2  (Mass consortium) & &  &  &\checkmark & &  \\
    
    NIST Tandem and Electron Ionization Mass spectral library, 2023 release  & & & & \checkmark & & \\
    
    LipidSearch (Thermofisher) & & \checkmark & \checkmark &  & & \\
    
   Wiley Registry\textsuperscript{\textregistered} of Mass Spectral Data 2023 & & & &\checkmark  & &\\
    
    Wiley Registry\textsuperscript{\textregistered} of  Tandem Mass Spectral Data - MS for ID  & & & & \checkmark &  &\\
    \hline
\end{longtblr}
       \SetCell[c=7, r = 1]{l}
      \textsuperscript{\hyperref[fn:tab2_one_top]{1}}\hspace*{0.05cm} For academic and non-commercial use.\protect\phantomsection\label{fn:tab2_one_bottom} \textsuperscript{\hyperref[fn:tab2_two_top]{2}}\hspace*{0.05cm} Download page contains non-redundant mass spectra that were calculated from available multiple replicate spectra. \protect\phantomsection\label{fn:tab2_two_bottom}  \textsuperscript{\hyperref[fn:tab2_three_top]{3}}\hspace*{0.05cm} Provides a tool to make your own database with computational data. \protect\phantomsection\label{fn:tab2_three_bottom}\textsuperscript{\hyperref[fn:tab2_four_top]{4}}\hspace*{0.05cm} Stores spectra of compounds tentatively identified. \protect\phantomsection\label{fn:tab2_four_bottom} 
\endgroup

Mass spectrometry data pipelines and infrastructures  are important to further grow mass spectral databases and to facilitate data management,  curation and  reproducibility \cite{Mendez2019}.  For example, Pedrioli and colleagues developed the open, vendor-independent data representation \texttt{mzXML}  in 2004, which enables cross-platform data analysis and management \cite{Pedrioli2004}. In addition, a plethora of freely available software has been developed to facilitate mass spectrometry data processing and upload, such as  OpenMS,\cite{rost2016openms} TidyMass,\cite{Shen2014} XCMS,\cite{Smith2006, Tautenhahn2008} metaboscape,\cite{metaboscape} progenesis,\cite{progenesis} mztab-m,\cite{Hoffmann2019} mzMine,\cite{Schmid2023} and MS-DIAL.\cite{Tsugawa2015} Furthermore, the GNPS database offers a feature-based molecular networking tool which connects feature processing to molecular network modeling. \cite{Nothias2020} 

Another important data management feature mitigates provenance variability. In LC-MS/MS mass spectrometry (as in other soft ionization techniques), data collected at different experimental conditions can vary in appearance. To mitigate such spectral variability, certain database providers have developed the concept of spectral trees\cite{Smith2005} and merged spectra\cite{Horai2010} that combine spectra collected under different conditions for the same analyte. 

\section{Compound identification: approaches and software}\label{sec:tools}
Compound identification is the primary purpose of mass spectral databases. Traditionally, compounds were identified by searching libraries or databases for matches. With the emergence of digital mass spectral databases more sophisticated approaches were developed, such as in silico fragmentation,\cite{Wolf2010,Ruttkies2016,Allen2014,Allen2015,Feunang2019} fragmentation trees,\cite{Durkop15, Durkop15fragtrees,Shen2014,Ludwig2020sirius} and machine learning approaches.\cite{Heinonen2012,Durkop15,Brouard2016,Nguyen2018, Nguyen2019}

In the traditional library search, the measured mass spectrum is compared to all spectra in a mass spectral database. The compound is identified (be it correctly or not) as the one with the most similar mass spectrum, out of those in the database. A mass spectral library search is inherently limited by the size of the database, which typically is some orders of magnitude smaller than the target compound space.\cite{Nguyen2019rev}

State-of-the art compound identification methods also use database information, but go significantly beyond library searches. Classical rule-based in silico fragmentation methods rely on a pre-defined set of chemical bond fragmentation rules to predict mass spectra,\cite{Nguyen2019rev} while combinatorial in silico fragmentation methods search all  possible fragmentation paths.\cite{Wolf2010,Ruttkies2016,Allen2014,Allen2015} During compound identification, spectra predictions are made for all entries in a compound database and compared to the measured spectrum to find the best match. In contrast to traditional mass spectral library searches, in silico fragmentation methods search through compound databases (e.g., PubChem) and not through mass spectral libraries. Compound databases cover a larger portion of chemical space than mass spectral databases, and are thus less limited in content and size. Rule-based in silico fragmentation methods are limited by the available fragmentation models that rely on heuristic bond energies (measured or estimated), while combinatorial methods generally need to limit the amount of fragmentation allowed by the model. In a similar vein, fragmentation tree methods find the optimal fragmentation tree that matches a recorded spectrum. Fragmentation trees are used for de novo molecular formula annotation through Gibbs sampling and Bayesian statistics.\cite{Ludwig2020zodiac,Ludwig2020sirius} In in silico fragmentation and fragmentation tree methods, machine learning is not necessarily a component, but can be included (e.g., competitive fragmentation modeling (CFM) method).\cite{Allen2014,Allen2015,Feunang2019}

\begin{figure}[H]
  \includegraphics{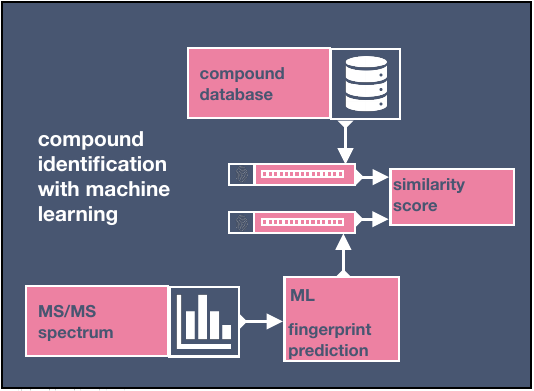}
  \caption{Schematic of the operating principle of most machine learning based compound identification tools. A machine learning model learns to map a mass spectrum to a feature space, here represented by a molecular fingerprint vector. In a second step, the similarity is scored between the predicted fingerprint and the molecular fingerprints of a compound database. Acronyms: ML - machine learning; MS/MS - tandem mass spectrometry.}
  \label{fig:mlms}
\end{figure}

The third category of compound identification algorithms is referred to as machine learning approaches, which are emerging as powerful property and structure inference tools in spectrometry.\cite{Kulik/etal:2022} \textbf{Figure \ref{fig:mlms}} illustrates the working principle of most compound identification machine learning algorithms.\cite{Heinonen2012,Durkop15,Brouard2016, Nguyen2018,Nguyen2019} In the first step, a mass spectrum is mapped to a feature space  represented by a so-called fingerprint. A fingerprint is a vector that encodes the presence or absence of certain molecular features, or their counts. Molecular fingerprints can be calculated in different ways from a molecular representation, like a 2D-molecular geometry (e.g., refs \cite{James1995,Landrum2022}). The mapping from spectra to molecular fingerprints requires a reference dataset of spectrum-molecule pairs. Supervised machine learning algorithms are then trained to assign fingerprints to spectra. Examples include  kernel methods, such as support vector machines,\cite{Heinonen2012} vector valued kernel ridge regression,\cite{Brouard2016, brouard2017, Brouard2019} and multiple kernel learning support vector machines,\cite{Shen2014, Durkop15, Durkop2019, Nguyen2018} or a combination of deep learning and multiple kernel learning.\cite{Nguyen2019} In the second step, the fingerprint vector is compared to the molecular fingerprints of compounds in compound databases. Moreover, compounds not present in any database can be annotated through hybrid searches.\cite{Durkop2020,Stravs2022,Cooper2019, Wallace2023} Additional information channels such as LC retention times,\cite{Stravs2022,Qiu2018,Gupta2011,Miller2013,Heravi01} pairwise retention orders \cite{witting2020current} or retention indices\cite{Qiu2018,Gupta2011,Miller2013,Heravi01,Stravs2022} (both relating to the retention order of compounds from LC), or collision cross sections\cite{Plante2019} can further improve the identification success. For retention time data, the heterogeneity of data across different laboratories is a hindrance, because the retention times depend on the configuration of the chromatograph. Machine learning techniques have been developed to standardize retention times across different laboratories\cite{stanstrup2015predret} and learn from the relative retention times of molecules,\cite{Bach2018,Bach2022} which are known to be more invariant across laboratories than absolute retention times.\cite{witting2020current}

Open access mass spectral databases containing high quality reference mass spectra have been essential for the development of machine learning based compound identification. For example, FingerID,\cite{Heinonen2012}  IOKR,\cite{Brouard2016} Adaptive,\cite{Nguyen2019} CSI:FingerID 1.0,\cite{Durkop15}  and CSI:FingerID 1.1\cite{Durkop2019} were all trained using different sets of compounds from different libraries (MassBank, GNPS, MassHunter Forensics/Toxicology PCDL library (Agilent technologies, Inc) and NIST17), with sizes ranging from approximately 1200 to 16083 compounds. The increase in compound identification accuracy during the past decade can largely be attributed to the growth of the spectral databases. In these examples, Agilent technologies, Inc and the NIST mass spectral library are the only commercial datasets.

In summary, a variety of approaches and software are now available for compound identification.  Open access mass spectral databases have been integral to the development of machine learning approaches and have facilitated the emergence of data-driven mass spectrometry in metabolomics. We will review in the next section how this insight, concepts, tools and infrastructures can be transferred to atmospheric science.

\section{Towards data-driven compound identification in atmospheric mass spectrometry}\label{s:ASDB_use}

In principle, all compound identification approaches we reviewed in this perspective could be directly used in atmospheric science. Suitable training or reference data, however, might be a limiting factor. The identification success rate would strongly depend on the number of atmospheric compounds in available mass spectral databases, or at least on the similarity between these compounds and those in the databases. Furthermore, the preferred mass spectrometric techniques in atmospheric science may differ from those prevalent in current databases. While compound identification algorithms may be able to extrapolate to the chemical space of atmospheric compounds, such generalization would be algorithm dependent and likely incur large uncertainties. We will address these points and propose an action plan to improve data-driven compound identification in atmospheric science. We start off by highlighting general challenges faced in the adoption of mass spectral databases for data-driven compound identification. 

\subsection{Data heterogeneity in mass spectrometry databases}
The content coverage of current mass spectrometry databases is heterogeneous in terms of compounds, instruments and experimental procedures. Tool and method developers therefore face the challenge of balancing the available data volume, more of which is beneficial for, e.g., machine learning methods, against the increased effort of handling the heterogeneity appropriately. Another challenge is the aforementioned coverage overlap, which could introduce biases in data-driven tools derived from more than one database. The current extent of this overlap, is, unfortunately, not known, since the last investigation by Vinaixa et al. dates back to 2016.\cite{Vinaixa2016} The heterogeneity of available mass spectrometry techniques (see Figure \ref{fig:pies_on_massbanks}b) presents a further challenge, but also an opportunity. The characteristics of spectra produced by different mass spectrometry techniques differ, which necessitates dedicated tool and method development. In the long run, however, this technique diversity could be advantageous since different spectrometries could complement each other synergistically. With transfer learning, multivariate machine learning models could be trained to convert between techniques or operate directly on heterogenous datasets. 


In summary, in atmospheric science much work is still required to assess the utility of existing databases, determine which training data to include in new models, and to establish initial identification tools for atmospherically relevant compounds. Below we provide a first assessment of the relevance of current mass spectral libraries for data-driven atmospheric mass spectrometry. Investments in improved compound identification for atmospheric science can be justified by the progress achieved in other application domains, such as metabolomics, which have been able to collect experimental data for tens of thousands of compounds (see Section \ref{sec:mb_tools}).

\subsection{Compound coverage of atmospheric molecules}\label{sec:compound_coverage}
As alluded to in Section \ref{sec:mb_tools}, atmospheric compounds are currently underrepresented in mass spectral databases. Compound identification approaches, that were developed for specific database compounds, will almost certainly perform worse for atmospheric compounds, than for compound classes in the databases.  This is true for  traditional library searches, which can only identify structures stored in a mass spectral database, as well as for algorithms built with database compounds and spectra.  

\begin{figure}[H]
  \includegraphics{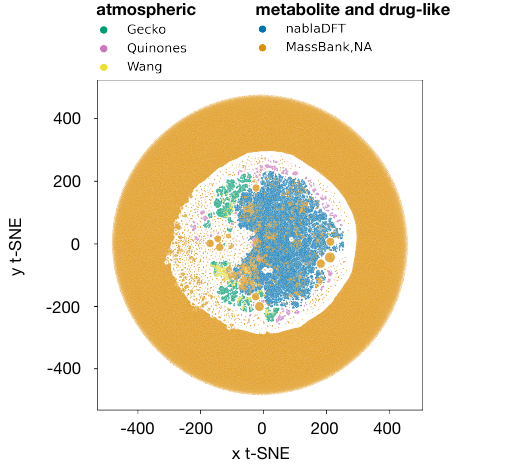}
  \caption{Similarity between molecular datasets containing drug molecules (nablaDFT), metabolites (Massbank of North America) and atmospheric molecules (Gecko, Wang and quinones) shown through  t-SNE clustering. The molecules were compared based on their topological fingerprint.}
  \label{fig:similarity}
\end{figure}

How well compound identification algorithms perform for atmospheric compounds depends on the overlap of atmospheric compound space with available mass spectral databases. \textbf{Figure \ref{fig:similarity}} shows a first visualization of this overlap. The figure presents a t-stochastic neighbourhood embedding (t-SNE) analysis for three atmospheric molecular datasets (here referred to as Gecko,\cite{Isaac2021,Besel/etal:2023} Wang\cite{Wang2017} and Quinones\cite{Kruger22, Tabor2019}) and two datasets of drug and metabolite compounds, representative of those in mass spectral databases (nablaDFT\cite{Khrabrov2022,Polykovskiy2020} and Massbank of North America\cite{MONA23}). t-SNE  clustered the compounds according to the similarity of their (molecular) topological fingerprints.\cite{James1995,Landrum2022} Figure \ref{fig:similarity} shows that the atmospheric compounds cluster closer together and are therefore more similar. Their clusters do, however, not overlap strongly, which indicates that these three datasets cover different parts of atmospheric compound space. The drug and metabolite compounds form their own clusters, most notable the dense ring of MassBank molecules surrounding the clusters of the other datasets. The two drug and metabolite datasets share some similarity in the inside of the ring, but only the MassBank has some small overlap with the three atmospheric datasets. The implications of Figure  \ref{fig:similarity} are: i) most atmospheric compound classes are absent from mass spectral databases;  ii) most atmospheric compounds therefore belong to a chemical space unknown by current compound identification algorithms; iii) the performance of compound identification algorithms in atmospheric science is unpredictable. Three traditional library searches report identification rates of only 2-35\% for atmospheric molecules,\cite{Hamilton2004,Worton2017,Franklin2022} providing further evidence for our three suppositions.  

Having established that atmospheric compounds differ from those in available mass spectral databases implies that compound identification algorithms would have to be able to extrapolate to be applicable in atmospheric science in the short term. Yet,  classical rule-based in silico fragmentation algorithms generalize poorly due to built-in rule-sets for chemical bond fragmentation,\cite{Nguyen2019rev} while in silico fragmentation methods based on combinatorial search (e.g. MetFrag, CFM-ID) are expected to do slightly better. On the other hand, generalization is a common challenge for machine learning models in chemistry.\cite{Mervin2021} For example, a machine learning model is forced to generalize when it evaluates a new elemental composition,\cite{Hu2022} molecular size\cite{Ghose2023} or functional group\cite{Scalia2020} that was not in the training data. Methods for quantifying uncertainty or confidence in a model's prediction have been developed through ensemble methods,\cite{Ghose2023, Wan2021} Bayesian neural networks,\cite{Zhou2023} Gaussian Process regression,\cite{Fang2021} support vector machines,\cite{hoffmann2022high} and Monte Carlo dropout.\cite{Wen2020} In metabolomics, it has been shown that machine learning methods predicting molecular fingerprints from spectra out-perform in silico fragmentation approaches.\cite{Durkop2019,Bach2022} However, it is not known if this also holds true in atmospheric science, where the coverage of the reference spectra of the relevant chemical space is significantly smaller.

Until atmospheric data is available in large enough quantities in mass spectral databases, it would seem prudent to not develop new compound identification methods or workflows immediately for atmospheric science. Machine learning-based approaches, for example, could instead evolve from existing methods developed in other application domains by means of transfer learning. For mass spectrometric techniques commonly found in mass spectral databases, such as tandem mass spectrometry or EI-MS, transfer learning would be particularly well-suited, as already developed models would likely only have to be retrained on  atmospheric data. However, for underrepresented techniques such as Api-CIMS, transfer learning would not be applicable and new approaches would have to be developed. Api-CIMS applications are currently flourishing in atmospheric science (see Section \ref{sec:MSvol}),\cite{LopezHilfiker14,Rissanen2019,Riva2019,Brophy2015,Skytta2022,Partovi2023,Krechmer2016,Berndt2018,He2023} but are practically absent from current databases (e.g., less than 0.1\% of the European MassBank\cite{MassBankEu22} data, see Figure \ref{fig:pies_on_massbanks}b). If atmospheric science is moving towards data-driven compound identification, this severe lack of data needs to be addressed. In the following we outline an action plan to fill this data vacuum. 

\subsection{Action plan}
\begin{figure}[H]
  \includegraphics[width=0.3\columnwidth]{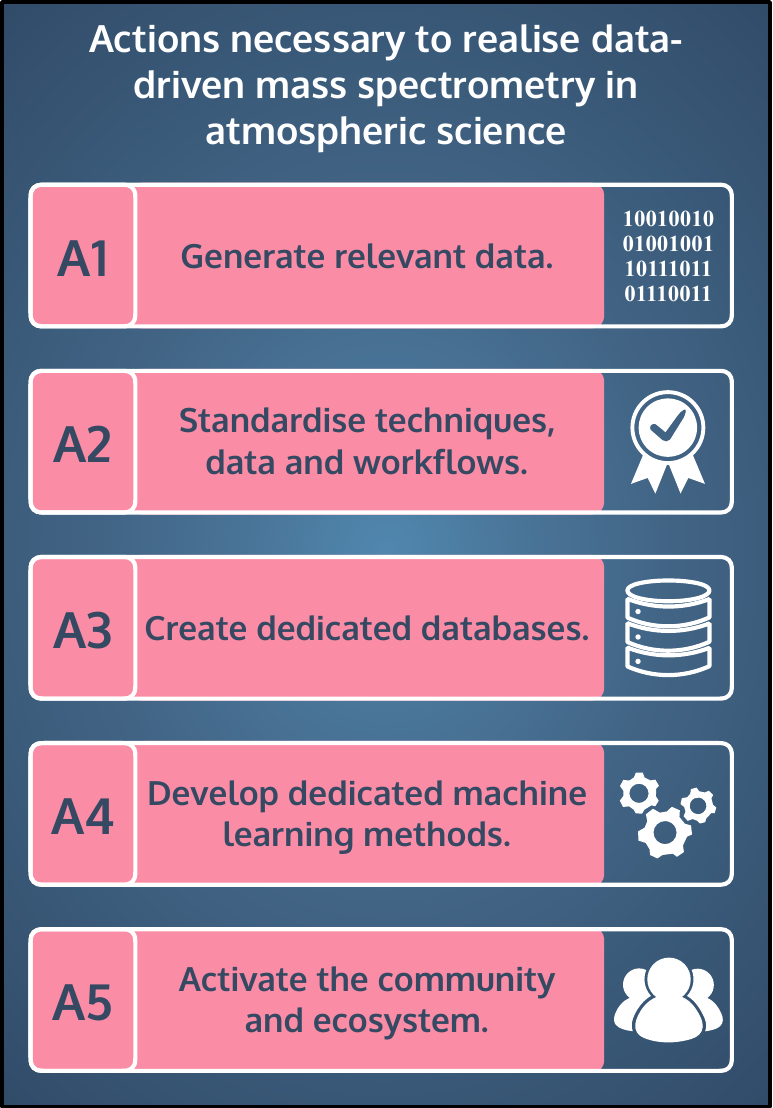}
  \caption{Our proposed action plan is designed to overcome the challenges hindering an successful implementation of data-driven mass spectrometry in atmospheric science. The plan contains five steps A1-A5.}  
  \label{fig:AP}
\end{figure}

In this perspective we reviewed the current challenges of implementing data-driven methods for mass spectrometry in atmospheric science. We next present practical strategies to overcome the identified barriers. Our recommendations are summarized in Figure \ref{fig:AP} and expanded on in the following.
 
\subsubsection{A1 -- Relevant data.} A paradigm shift towards data-driven mass spectrometry in atmospheric science could begin with access to relevant data (Section \ref{s:ASDB_use}). For atmospheric mass spectrometry, reference spectra would have to be collected for the compounds taking part in atmospheric chemistry, including the atmospheric gas-phase, small clusters and nanoparticles (see Section \ref{sec:MSvol}). The collection could begin with representative compounds and expand from there. Finding such relevant molecules is no simple feat, because the chemical space of atmospheric compounds is large and largely uncharted. We suggest to use data-driven approaches, possibly based on the volatility basis set description of atmospheric compound space  (see  Section \ref{sec:intro}),  to ensure data collection of compounds with varying properties of interest, such as, e.g., volatility and O:C ratio.  Data collection should furthermore include the multiple mass spectrometry techniques used in atmospheric science for compatibility with existing databases and compound identification tools, as well as for a holistic description of atmospheric chemistry. It is particularly important to include presently underrepresented techniques (e.g., Api-CIMS, as addressed in Section \ref{sec:compound_coverage}) to improve their data coverage in the databases. The methodology portfolio could be augmented with synthetic data generated with computational tools as discussed further in A4 below. For example, computational studies in atmospheric chemistry have shown that the binding energy between  molecules and reagent ions can be used to predict the experimentally measured CIMS sensitivity  e.g., refs \cite{Iyer2016,Hyttinen2017,Hyttinen2018}.  

\subsubsection{A2 -- Standardization.} To  utilize the collected data in atmospheric science to its full extent, standards and standardized practices for data collection, curation, management and sharing need to be agreed on and implemented. For certain mass spectrometric techniques (e.g., EI-MS and MS/MS),  such practices have already been developed in other fields (e.g., metabolomics, see Section \ref{sec:mb_tools}) to ensure data standardization and reproducibility (for example, platform-independent data formats, data analysis pipelines and spectral trees or merged spectra). They could be directly applied to atmospheric mass spectral data and should be embraced by atmospheric scientists. Conversely, for techniques currently underrepresented in mass spectral databases (e.g., Api-CIMS), appropriate standardization practices still need to be developed. Such practices also need to consider the specific use-cases in atmospheric science (e.g., the lack of sample pre-treatment and separation by chromatography). For example, Api-CIMS data should be easy to standardize, because the number of different Api-CIMS instruments used in the field has stayed relatively small, with a dominant fraction of the data being acquired by similar methods, such as chemical ionization atmospheric interface time-of-flight (CI-Api-ToF) instrumentation, or the recently introduced orbitrap CIMS systems.\cite{Junninen2010,Kurten11,Jokinen2012,Riva2019,Riva2020} For Api-CIMS, the standardization of ion production and gas-phase sample introduction is crucial for ensuring fully reproducible measurements. The signal depends on specific ion-molecule reactions and interaction time. Gas-phase chemical ionization is typically linear and scalable, allowing for a wide range of ion concentrations for increased sensitivity. Normalizing measured signals with the number of charge carriers (i.e., reagent ions) is essential in Api-CIMS analysis to account for differences in the initial ion pool. Digital CI-Api-ToF twins can aid in the standardization.\cite{Passananti/etal:2019}

\subsubsection{A3 -- Infrastructure.} Data collection and sharing require dedicated infrastructures. En route towards data-driven science, atmospheric science could proceed in two different ways: i) establish dedicated mass spectral databases for atmospheric science data that are operated by the atmospheric science community, or ii) contribute  atmospheric science data to existing mass spectral databases. A dedicated database in option i) offers better control over the data (for example, data curation, labeling and quality control), but requires concerted actions of key stakeholders and sustained funding.\cite{Himanen/Geurts/Foster/Rinke:2019}  Adopting existing mass spectral databases as in option ii), is therefore easier in the short term. Contributing to an existing, interdisciplinary mass spectral database promotes data sharing with the broader mass spectrometry community, which expands the user base. We recommend  a third option, which is an amalgamation of the two approaches above: curating dedicated databases, that can be local to research groups or consortia, but are regularly uploaded and synchronized with large open access databases (such as the MassBanks or GNPS). Dedicated databases could, for example, be linked to collections of reference spectra of atmospheric compounds (e.g., refs \cite{Claeys2007,Parshintsev2008,Lin2012,eijck2013}). Such collections need to grow to provide access to curated high-quality training data for the data-driven method development. Meanwhile, data from field campaign repositories containing data of unknown compounds can be shared for compound identification. In addition, community datasets, such as refs \cite{Worton2017,Yee2018,Jen2019,Besel/etal:2023}, could complement data infrastructures. They offer distinct advantages such as  having been  purposefully curated with design criteria like similarity and balance in mind.

\subsubsection{A4 -- Dedicated machine learning methods.} 
\begin{figure}[H]
  \includegraphics{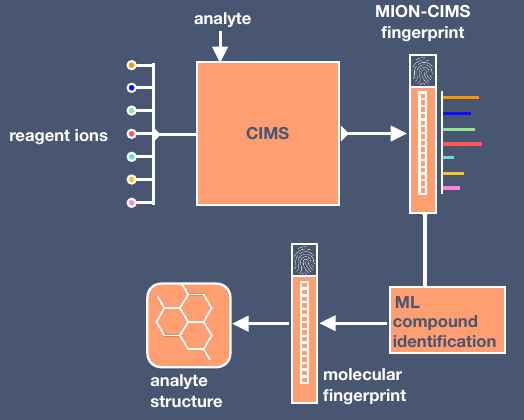}
  \caption{A proposed workflow for machine learning based compound identification with MION-CIMS. The model learns how to map the molecule specific MION-CIMS fingerprint (set of CIMS sensitivity values for different reagent ions) to a molecular representation.}
  \label{fig:cimsml}
\end{figure}

In Sections \ref{sec:tools} and \ref{sec:compound_coverage}, we reviewed the potential and challenges of available machine learning-based compound identification tools in atmospheric science and observed that the identification performance depends strongly on the availability of relevant data (see A1). For tandem and EI-mass spectrometry, data is available for other compounds and  we propose to begin applying existing machine learning techniques to atmospheric data and to then refine the models accordingly. Over time, such models could be improved through transfer-learning, possibly coupled to active learning schemes, as new atmospheric data becomes available (Section  \ref{sec:compound_coverage}).  For mass spectrometric techniques, which lack existing machine learning models, but are used for compound identification in atmospheric science (e.g.,  MION-CIMS), new, dedicated models need to be developed. \textbf{Figure \ref{fig:cimsml}} outlines our proposal for a machine learning-based compound identification scheme for MION-CIMS. The CIMS  sensitivity for different reagent ions acts as the molecule-specific MION-CIMS fingerprint. The machine learning model learns how to map the MION-CIMS fingerprint to a molecular representation. The development of such a new machine learning-based model could make use of computational mass spectral databases until experimental counterparts become available (see A1). To that end, machine learning could also assist in building computational databases by expediting calculations of the binding energies used to predict CIMS sensitivity.

\subsubsection{A5 -- Community endorsement.} Wide-spread adoption of standardised data practices requires a community wide effort. Together, the atmospheric science community needs to commit to open data sharing and publishing. The data should preferably be shared through open access databases, or with FAIR sharing rights,\cite{Wilkinson2016} if published with commercial parties. Adoption of community wide-data practices can be encouraged through education in data literacy and machine learning, for example in summer schools, webinars or workshops. Further dissemination at atmospheric science conferences and through research networks would create awareness and rally the community to endorse the new paradigm.

\section{Take-home message}
In this perspective, we reviewed the current state and potential for data-driven compound identification in atmospheric mass spectrometry. Although developments of experimental techniques now enable monitoring and tracking of atmospheric chemical processes, an accurate method for high-throughput compound identification is still missing. Community-wide efforts to improve data standardization and collection can support the transition towards reliable identification of atmospheric compounds with mass spectrometry. Integration of data-driven approaches, such as machine learning, into mass spectrometric data analysis will facilitate knowledge gain. Concomitantly, a true paradigm change requires a community endorsement and a combined effort to collect, curate and share data in a standardized manner. Although the development of data-driven approaches requires an initial time and resource investment,  data-driven approaches promise to be more efficient than the manual processing currently employed. Successful examples in parallel fields can be used to guide and inform this shift towards a digital era in atmospheric mass spectrometry. 

\medskip

\medskip
\textbf{Acknowledgements} \par 
This project has received funding from the European Research Council under the European Union’s Horizon 2020 research and innovation programme under Grant No. 101002728. The support from the Academy of Finland (334790, 339421, 345802, 346373, 346377, 353836, 358066) is greatly appreciated.

\medskip

%


\newpage
\begin{figure}
\textbf{Table of Contents}\\
\medskip
  \includegraphics{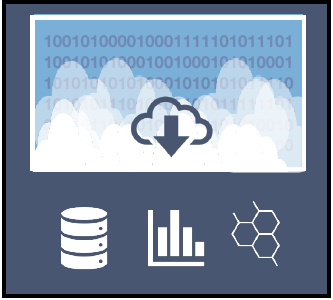}
  \medskip
  \caption*{ToC Entry}
\end{figure}

\end{document}